# Engineering on-surface spin crossover: spin-state switching in a self-assembled film of vacuum sublimable functional molecule

Kuppusamy Senthil Kumar,[1*] Michał Studniarek,[1,2] Benoît Heinrich,[1] Jacek Arabski,[1] Guy Schmerber,[1] Martin Bowen,[1] Samy Boukari,[1] Eric Beaurepaire,[1] Jan Dreiser,[2] Mario Ruben[3*]

[1] *Institut de Physique et Chimie des Matériaux (IPCMS), Université de Strasbourg, F-67034 Strasbourg, France,*
\* e-mail : senthil.kuppusamy@ipcms.unistra.fr

[2] *Swiss Light Source, Paul Scherrer Institut (PSI), CH-5232 Villigen, Switzerland,*

[3] *Institute of Nanotechnology (INT), Karlsruhe Institute of Technology (KIT), Hermann-von-Helmholtz-Platz 1, D-76344, Eggenstein-Leopoldshafen, Germany.*
\* email: mario.ruben@kit.edu

**Abstract.** Realization of spin crossover (SCO) based applications requires studying of spin state switching characteristics of SCO complex molecules at nanostructured environments especially on-surface. Except for a very few cases, the SCO of a surface bound thin molecular film is either quenched or heavily altered due to (i) strong molecule-surface interactions and (ii) differing intermolecular interactions in films relative to the bulk. By fabricating SCO complexes on a weakly interacting surface such as highly oriented pyrolytic graphite (HOPG) and copper nitride (CuN), the interfacial quenching problem has been tackled. However, engineering intermolecular interactions in thin SCO active films is rather difficult. This work proposes a *molecular self-assembly strategy* to fabricate thin spin switchable surface bound films with *programmable intermolecular interactions*. Molecular engineering of the parent complex system [Fe(H$_2$B(pz)$_2$)$_2$(bpy)] (pz = pyrazole, bpy = 2,2'-bipyridine) with a dodecyl (C$_{12}$) alkyl chain yielded a classical amphiphile-like functional and vacuum sublimable charge neutral Fe$^{II}$ complex, [Fe(H$_2$B(pz)$_2$)$_2$(C$_{12}$-bpy)] (C$_{12}$-bpy = dodecyl[2,2'-bipyridine]-5-carboxylate). The bulk powder and 10 nm thin film, on quartz glass/SiO$_x$ surface, of the complex showed comparable spin state switching characteristics mediated by similar lamellar bilayer like self-assembly/molecular interactions in both bulk and thin film states. This unprecedented observation augurs well for the development of SCO based applications, especially in molecular spintronics.

Metal-organic spin crossover (SCO) complexes[1–4] capable of undergoing switching between low-spin (LS) and high-spin (HS) states as a function of temperature, light, pressure, and electric field are suitable candidates to fabricate room temperature operable molecular electronic/spintronics architectures.[5–10] Relentless efforts have been made to study spin state switching characteristics of SCO complexes on different surfaces[11–19] to harness the device utility of SCO entities and several interesting results, e.g., the spin-state dependence of electrical conductance,[6,20] memristance behaviour,[21] electric field or electron induced SCO,[22,23] and locking and unlocking of SCO around room temperature[24] have been reported. Vacuum sublimation of the complexes on to a suitable surface is the preferred method to obtain high-quality surface layers of SCO complexes.[25–30] Despite continuing efforts only a few families of sublimable SCO complexes have been realized, these include charge neutral [Fe(phen)$_2$)(NCS)$_2$] (phen = 1,10-phenanthroline),[31] [Fe(H$_2$B(pz)$_2$)$_2$(L)] (pz = pyrazole, L = 2, 2'–bipyridine (bpy) or 1,10-phenanthroline (phen)),[32,33] and [Fe(HB(pz)$_3$)$_2$][34] systems. Among them, the [Fe(H$_2$B(pz)$_2$)$_2$(L)] complexes have been studied in detail on metallic Au(111),[18,35–37] non-metallic Bi(111),[13] and highly oriented pyrolytic graphite (HOPG) surfaces.[17] The switching is reported to be coverage dependent on metallic Au(111) substrate on which the first layer of the complex undergoes fragmentation and loss of SCO,[38] whereas switching is retained on Bi(111) and HOPG surfaces even at sub-monolayer level due to the weak molecule-surface interfacial interactions.[13,17] While the role of the surface in affecting spin state switching of SCO thin film architectures is well established, control of the intermolecular interactions, a governing factor in dictating SCO in the bulk phase, is a rather difficult problem to tackle at nanostructured environments especially on surface bound thin films due to lack of control associated with the film fabrication processes in engineering intermolecular interactions warranting systematic experimentation in this direction. To this end, a self-organization strategy using tailored intermolecular interactions could be employed. This could be realized by the synthesis of a *functional-SCO complex* with a propensity to self-assemble via intermolecular interactions, which could be copied *"as such"* when translating from bulk to thin film state enabling the realization of similar if not identical SCO characteristics between bulk and thin film phases. To start anew in this direction, a functional [Fe(H$_2$B(pz)$_2$)$_2$(C$_{12}$-bpy)] (C$_{12}$-bpy = dodecyl [2,2'-bipyridine]-5-carboxylate) SCO complex (cf. Figure 1a), whose structure is reminiscent of classical amphiphile, featuring SCO active pseudo-hydrophilic head and hydrophobic dodecyl (C$_{12}$) tail groups, capable of self-assembly via intermolecular hydrophobic interactions has been designed and synthesized. Spin state switching characteristics of the [Fe(H$_2$B(pz)$_2$)$_2$(C$_{12}$-bpy)] complex in bulk and thin film state has been probed to elucidate the role of intermolecular interactions in mediating SCO characteristics upon moving from bulk to thin film environment.

Magnetic susceptibility measurement of [Fe(H$_2$B(pz)$_2$)$_2$(C$_{12}$-bpy)] complex powder showed gradual SCO with $T_{1/2}$ = 210 K (160 K for the parent complex) without any noticeable hysteresis upon repeated cycling (cf. Figure 1b). The obtained $\chi_m T$ product of 3.55 cm$^3$ K mol$^{-1}$ and 0.05 cm$^3$ K mol$^{-1}$ at 300 K and 100 K, respectively, indicate the presence of pure high and low spin species at those temperatures and genuine temperature-induced SCO behavior. To get insights into the SCO characteristics and self-assembly of the [Fe(H$_2$B(pz)$_2$)$_2$(C$_{12}$-bpy)] in the thin film state, the complex was sublimed on quartz glass substrate and SiO$_x$ surfaces from a crucible held at a temperature of 150 °C, far below the onset of degradation inferred from thermogravimetric analysis (TGA) (cf. Figure S1). Electronic spectral analysis of the thin film sample sublimed on quartz glass showed the ligand centred and metal to ligand charge transfer $^1$(MLCT) bands analogues to the chloroform solution of the complex (cf. Figure S2a and b) confirming structural integrity of [Fe(H$_2$B(pz)$_2$)$_2$(C$_{12}$-bpy)] on quartz substrate after sublimation.

The SCO characteristics of a 10 nm thick film of [Fe(H$_2$B(pz)$_2$)$_2$(C$_{12}$-bpy)] on a SiO$_x$ surface has been probed using X-ray absorption spectroscopy (XAS) at the Fe $L_{2,3}$ edges. Comparable XAS spectra recorded at room temperature for the 10 nm thick film on SiO$_x$ to the powder sample of the complex further confirms the conserved electronic structure of the complex in the film (cf. Figure 2). Upon cooling the bulk powder and the thin film of [Fe(H$_2$B(pz)$_2$)$_2$(C$_{12}$-bpy)] from 300 K to 100 K, a spectral intensity ratio change of Fe $L_3$ edge multiplet features at ~ 706.8 eV and ~ 708.1 eV typical for HS $t_{2g}^4 e_g^2$ configuration (S = 2) and LS $t_{2g}^6 e_g^0$ (S = 0), respectively, have been observed in line with the previous studies.[18] A decrease in the branching ratio,[18,39] defined as $b = I_3/(I_3+I_2)$, $I_x$ = integrated intensity at the corresponding $L_x$ edge, from $b = 0.7$ at 300 K to $b = 0.61$ at 100 K, indicative of reduced spin-orbit coupling due to the diamagnetic nature of LS state, further confirms the occurrence of SCO in the thin film sample. Variable temperature XAS measurements have been performed to get insights into the thermal SCO characteristics of bulk powder and 10 nm film of [Fe(H$_2$B(pz)$_2$)$_2$(C$_{12}$-bpy)]. By using the peak intensity ratios from the HS (LS) reference curve recorded at 300 K (100 K) the HS proportion of the complexes within the probed region at a given temperature has been assessed (cf. Figure S3c). The analysis established $T_{1/2} = 197$ K for the bulk sample, which is ca. 13 K lower than the one recorded for the powder sample using SQUID magnetometry and is most probably due to temperature calibration issues. The transition curve of the 10 nm film is clearly more gradual in character with transition temperature shifted by ~20 K towards lower values with respect to the bulk sample, this could be attributed to a weak effect of the interface or to the slightly different intermolecular interactions among the molecules in sublimed film in comparison with the powder form (vide infra).[40]

X-ray diffraction (XRD) measurements have been performed to investigate the nature of the self-organization of [Fe(H$_2$B(pz)$_2$)$_2$(C$_{12}$-bpy)] in the bulk and the thin film states. The powder pattern of the bulk sample is characteristic of a lamellar crystal phase, as it is composed of a series of strong and sharp fundamental and higher-order reflections of the layer sequence (reflections (00$l$), periodicity $d = 25.4$ Å), and by some weak and unresolved crossed reflections of the three-dimensional crystalline cell (cf. **Figure 3**a). This arrangement directly follows the "amphiphile-reminiscent" design of the molecules, as they bind covalently two antagonistic segments, namely the alkyl chains and the metal complex counterpart. These incompatible moieties spontaneously phase separate in juxtaposed domains and most often form multilayered lamellae as in this case.[41–43]

A side effect of the intercalation of different nature of layers is the loss of positional correlation between molecular segments of subsequent lamellae. The consequence is some extent of blurring out of the three-dimensional superstructure, as revealed by the lowly resolved crossed reflections. The pattern of the thin film sublimed on quartz glass only displays the (00$l$) reflections, presumably because the film grew oriented, with the lamellar system parallel to substrate. In this case, all the reflections other than (00$l$) lie out of the specular plane and thus beyond the measured reciprocal space zone. Remarkably, film and powder patterns exhibit identical lamellar periodicity and same intensity ratios in the reflection series. The lamellar structure of the bulk sample was therefore maintained in the film, which fulfills the experimental realization of an SCO complex maintaining the same phase in bulk powder and thin film states. Some differences might nevertheless exist in the in-plane arrangements and in the related three-dimensional structure, which could explain the gradual and lower $T_{1/2}$ temperature observed for the thin film of [Fe(H$_2$B(pz)$_2$)$_2$(C$_{12}$-bpy)] complex.

Beyond the information directly extracted from patterns, the common lamellar structure of bulk sample and film can be scrutinized with a geometrical analysis. The basis of this approach is the empirical rule that in nano-segregated systems, the geometrical features of the individual segments are widely independent of the molecular architectures associating them. These features can then be extracted from density and crystallographic data on reference molecules and compared with the molecular area $A_{mol}$ of the layer sequence, defined as the ratio of molecular volume $V_{mol}$ and lamellar thickness $d$.[44,45] In this way, the volume contribution of the SCO head $V_{SCO}$ was found to range between 620-660 Å$^3$, on the basis of the single crystal structure of the complex without the dodecyl carboxylate substituent[32] and of reference partial volume data. The volume contribution of the alkyl tail is either of 280-300 Å$^3$ for crystallized chains or ca. 320 Å$^3$ for molten chains. Hence, for the entire molecule, $V_{mol}$ ranges between 900-1000 Å$^3$, and $A_{mol}$, between 37-41 Å$^2$. This is about half of the transverse section of the SCO head ($\sigma_{SCO} \approx 80$ Å$^2$, also determined from the single-crystal data) and about twice the section of the stretched tail ($\sigma_{ch}$ = 18.5-20 Å$^2$ if crystallized, and $\sigma_{ch} \approx 21$ Å$^2$, if molten). The compacity of the whole multilayered structure is imposed by the requirement of the bulkiest segments, thus here the SCO heads arranged in double-layers to minimize $A_{mol}$ (cf. Figure 3b). The smaller tails need to spread for filling up the intervals between successive double-layers, which occurs through various easy mechanisms: in the crystallized state, tails generally arrange in strongly tilted double-layers, whereas molten tails rather fold or eventually interdigitate.[46] In any case, the amphiphile-like molecular organization of [Fe(H$_2$B(pz)$_2$)$_2$(C$_{12}$-bpy)] into double-layers spaced by tails is clearly established from XRD, as well as the identity of the lamellar structure between powder and thin film, and the spontaneous surface alignment of the lamellar system in the film. The morphology of the thin film was further investigated by tapping mode atomic force microscopy (AFM) experiments (cf. **Figure 4**) performed on 10 nm thick film of [Fe(H$_2$B(pz)$_2$)$_2$(C$_{12}$-bpy)] on SiO$_x$. Large-scale AFM image on Fig. 4a clearly reveals the formation of flat terraces by the molecules. A relative height of the steps either in range $d$ = 1.1 – 1.3 nm or $d$ = 2.2 – 2.6 nm was found in good agreement with the lamellar spacing deduced from XRD and corresponds to half or entire double-layer structures as deduced from the structural model in Fig. 3b.

The above results clearly indicate the role of intermolecular interactions in successfully preserving SCO in nanostructured thin films and signify the role of chemical structure alterations leading to desired material properties. Importantly, functionalization of the parent system with the dodecyl chain also resulted in a 50 K increase in $T_{1/2}$ value relative to its parent complex clearly elucidating the role of the chemical substituent in tuning SCO in line with the literature reports detailing alkyl chain tuning of SCO characteristics.[47–49] The parent [Fe(H$_2$B(pz)$_2$)$_2$(bpy)] complex was reported to grow on amorphous surfaces in a Volmer-Weber mode by crystallites formation, whereas functionalization of the [Fe(H$_2$B(pz)$_2$)$_2$(bpy)] head with the C$_{12}$ chain imposed a preferential stacking direction and facilitated vertically ordered growth on the surface beneficial in terms of fabricating multilayer vertical SCO junctions.[50] The alignment of [Fe(H$_2$B(pz)$_2$)$_2$(C$_{12}$-bpy)] perpendicular to the plane of the surface in thin films clearly indicates the strong intermolecular interactions overcoming molecule-surface interfacial interactions. This enables systematic distance scaling between the SCO head and surface via synthetic modification of alkyl chain length and could be considered as a novel strategy to decouple SCO heads from the surface, via insulating alkyl chains, like adding an insulating layer between the strongly interacting metallic surface and the SCO complex. Note also a recent report by Zhang et al.,[24] detailing quenching of SCO in 5 nm film of [Fe(H$_2$B(pz)$_2$)$_2$(bpy)] on SiO$_2$ and Al$_2$O$_3$; the preservation of SCO in 10 nm films observed in the present work is also important in this context and in terms of realizing large area spintronics junctions based on SCO complexes.

To summarize, the present study demonstrated a functionalization strategy of the parent [Fe(H$_2$B(pz)$_2$)$_2$(bpy)] complex system which alters the material properties by modifying the molecular stacking arrangement but preserved its spin transition and sublimation propensity. The functional [Fe(H$_2$B(pz)$_2$)$_2$(C$_{12}$-bpy)] complex could be sublimed at relatively low temperatures in comparison with its parent counterpart and was found to self-assemble in the same amphiphile-like lamellar structure in the bulk and the thin film state, while effecting near-identical switching characteristics of the complex in both states. To the best of our knowledge, this is the first successful attempt of sublimation of a *functional SCO complex* with the observation of SCO in a thin film environment. The results presented in this study are expected to open a way for a more molecular-engineering oriented approach in the SCO materials research direction and studying of on-surface switching characteristics of vacuum sublimable functional SCO complexes with strict control of the nanostructure morphology which has strong implications in determining charge carrier mobilities as reported for a poly(3-hexylthiophene) based lamellar self-assembled domains on SiO$_2$/Si substrate.[51] Further, several *functional addends* featuring advantageous physical properties such as luminescence and chirality could be coupled with the [Fe(H$_2$B(pz)$_2$)$_2$(bpy)] skeleton and the resultant functional-SCO complexes may be sublimed on surfaces which may result in upheaving of SCO based research direction towards realistic SCO based applications.

**Experimental Section**

Synthesis and characterization of the ligand and complex were detailed in the supporting information. The 10 nm thick film was sublimed from a molybdenum boat heated up by electric current to 120 ºC under pressure of $10^{-8}$ mbar using a Plassys deposition machine. The thickness of the film was established based on the quartz balance used during the process and calibrated with x-ray reflectivity performed on a test sample. The XAS measurements were performed at the X-Treme beam line at the Swiss Light Source, Paul Scherrer Institute, Switzerland.[52] The powder sample was pressed into an indium foil. The XAS spectra were recorded at normal X-ray incidence in total electron yield mode and normalized to the integral over the sum of $L_3$ and $L_2$ edges of Fe. Ultraviolet-visible absorption measurements were performed with a Varian Cary 5000 double-beam UV-vis-NIR spectrometer, with referential complex dissolved in chloroform ($CHCl_3$). Magnetic susceptibility measurements were performed on an MPMS-XL5 SQUID magnetometers (Quantum Design) at $B_{DC}$ = 0.1 T field and sweep rate 2 K min$^{-1}$. Gelatine capsules were used as sample holders in the temperature range 100 ↔ 300 K. The diamagnetic corrections of the molar magnetic susceptibilities were applied using Pascal's constants. The crystalline phase of the powder and the films were determined using a Bruker D8 Advance diffractometer equipped with a LynxEye™ detector in the θ - 2θ mode with a monochromatic wavelength $\lambda_{CuK\alpha 1}$ = 1.54056 Å), 0.2°/min over a 2θ range from 2 to 30° and at room temperature. Atomic force microscopy measurements of the complex sublimed onto $SiO_x$ surface were performed with Bruker Ikon microscope in tapping mode at ambient pressure and room temperature. The error bars of the height along the profile were established as a standard deviation of the points in flat regions above and below the step.

**Acknowledgements**

M. R. would like to thank the financial support of the Agence Nationale de la Recherche-Labex NIE 11-LABX-0058_NIE within the "investissement d'Avenir" program ANR-10-IDEX-0002-02 and Deutsche Forschungsgemeinschaft (DFG) through TRR 88 "3MET" (A1, A4, C4, C5, and C6). M. S. and J. D. gratefully acknowledge funding by the Swiss National Science Foundation (Grant no. 200021_165774/1). This project has received funding from the European Union's Horizon 2020 research and innovation programme under the Marie Skłodowska-Curie grant agreement No 701647. The authors also thank ANR-11-LABX-0058_NIE within the Investissement d'Avenir program ANR-10-IDEX-0002-02 for the AFM facility at IPCMS.

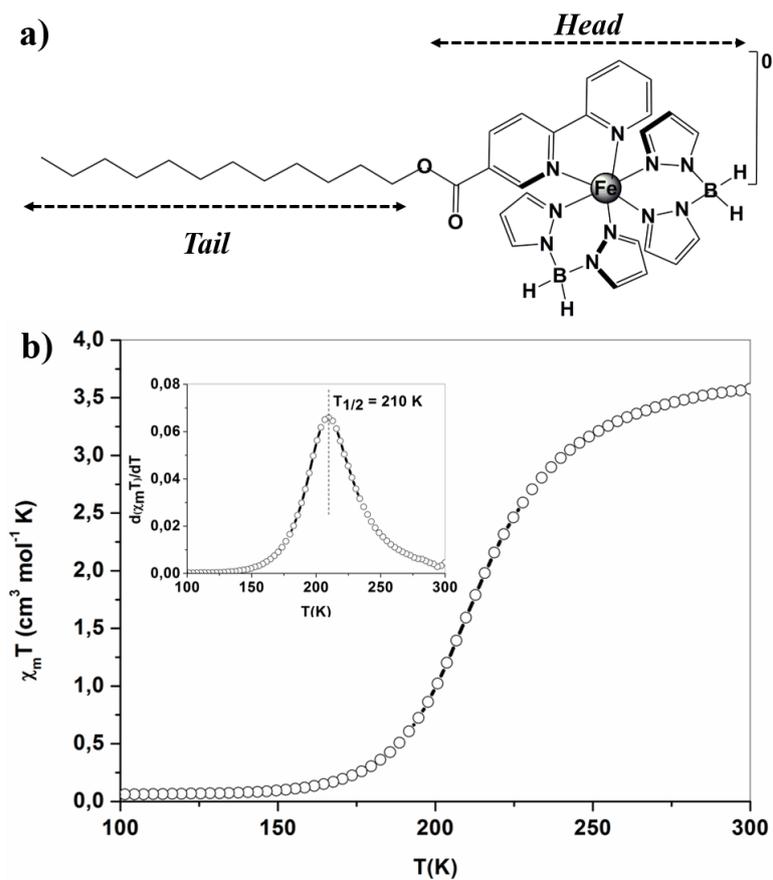

**Figure 1.** (a) Molecular structure of the pseudo-amphiphile like SCO complex [Fe(H$_2$B(pz)$_2$)$_2$(C$_{12}$-bpy)] and (b) $\chi_m T$ vs. $T$ plot of bulk powder form of [Fe(H$_2$B(pz)$_2$)$_2$(C$_{12}$-bpy)], inset shows d$(\chi_m T)$/d$T$ vs. $T$ curve indicating $T_{1/2}$ = 210 K.

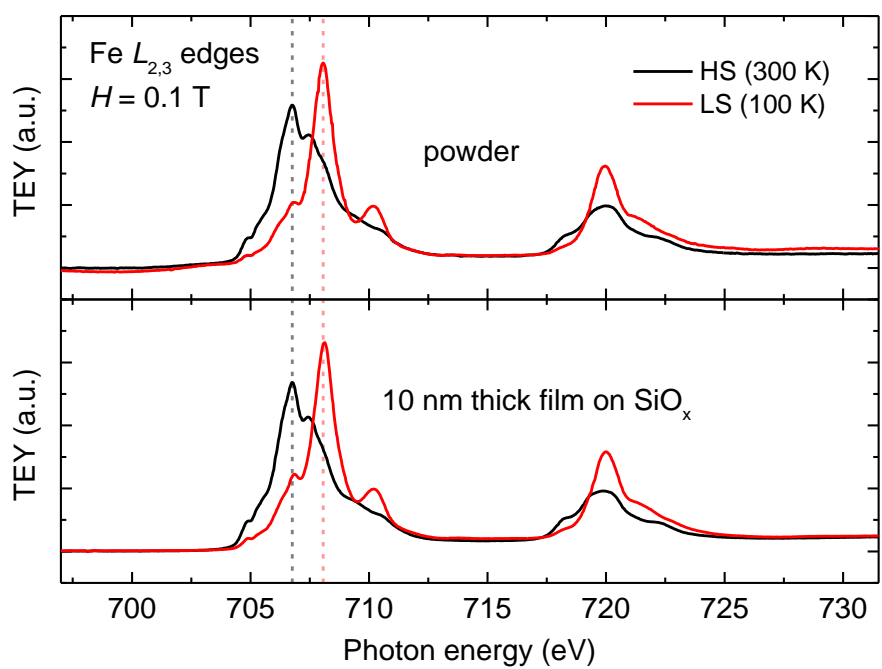

**Figure 2.** XAS at the Fe $L_{2,3}$ edges of [Fe(H$_2$B(pz)$_2$)$_2$(C$_{12}$-bpy)]. XAS spectra acquired for powder (top) and sublimed 10 nm thick film on SiO$_x$ (bottom) at 300 K and 100 K. The vertical dotted lines are guides for eyes indicating multiplet features at ~ 706.8 eV and ~ 708.1 eV characteristic for HS and LS state molecules, respectively, of bulk and thin film samples. The curves were recorded in total electron yield mode at normal X-ray incidence and normalized to the sum of integrals over the Fe $L_3$ and $L_2$ edges.

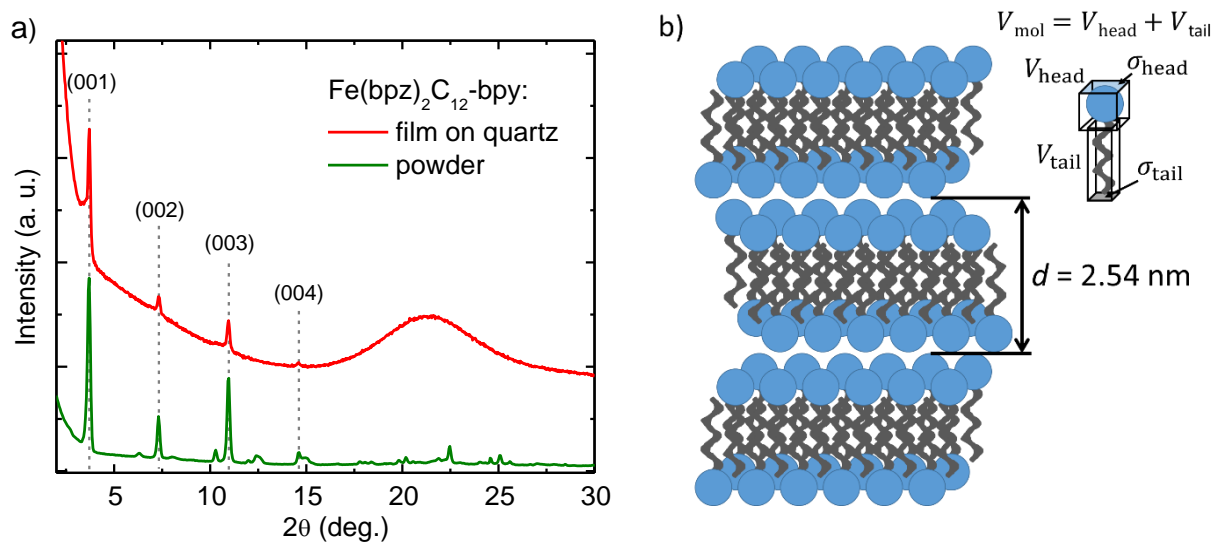

**Figure 3.** Self-organization of a [Fe(H$_2$B(pz)$_2$)$_2$(C$_{12}$-bpy)] thin film. (a) X-ray diffraction pattern of a 10 nm thick sublimed [Fe(H$_2$B(pz)$_2$)$_2$(C$_{12}$-bpy)] film on a quartz substrate compared to the powder reference. The equidistant reflections (001) – (004) reveal the lamellar structure of 2.54 nm periodicity in powder and film. The broad feature at 17 – 25° is the scattering signal from the quartz glass substrate. (b) Schematic view of the internal structure of lamellae of [Fe(H$_2$B(pz)$_2$)$_2$(C$_{12}$-bpy)] consisting of double-layers of the complex heads (blue spheres) alternating with aliphatic tails arranged in double layers, according to the deduced bilayer crystalline arrangement.

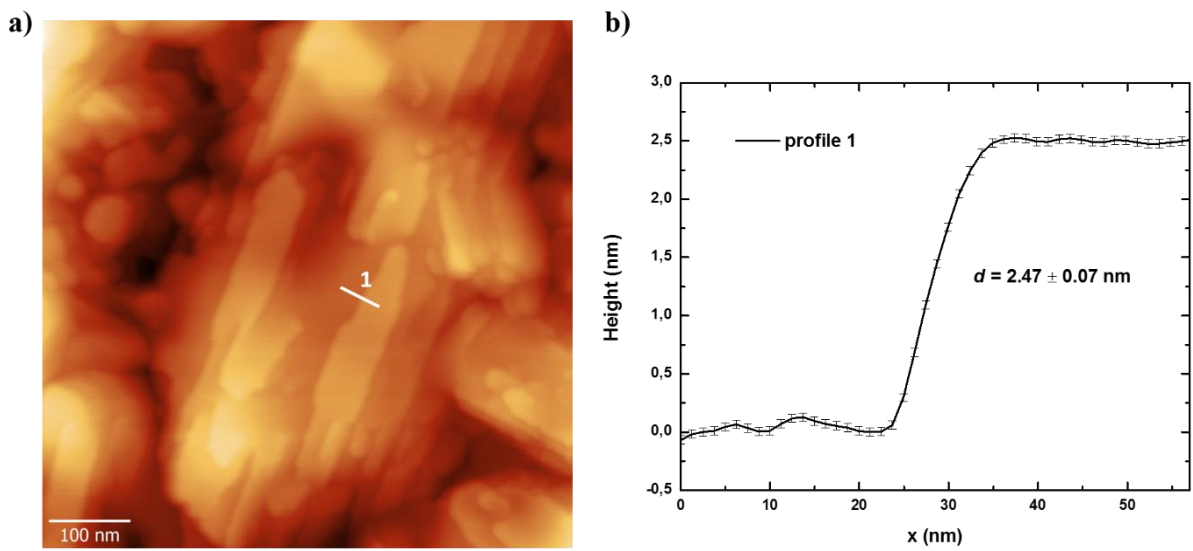

**Figure 4.** Surface morphology of a sublimed [Fe(H$_2$B(pz)$_2$)$_2$(C$_{12}$-bpy)] film on SiO$_x$ substrate. (a) Surface topography measurement performed by room temperature ambient pressure AFM in tapping mode of Si/SiO$_x$(400 nm)/[Fe(H$_2$B(pz)$_2$)$_2$(C$_{12}$-bpy)](10 nm) highlighting bilayers of the lamellar molecular stacking. (b) Height profile of the step along the profile 1 shown in (a) of height $d = 2.47 \pm 0.07$ nm consistent with lamellar spacing extracted from XRD.

# Supporting Information

**Synthesis of [Fe(H₂B(pz)₂)₂(C₁₂-bpy)]**

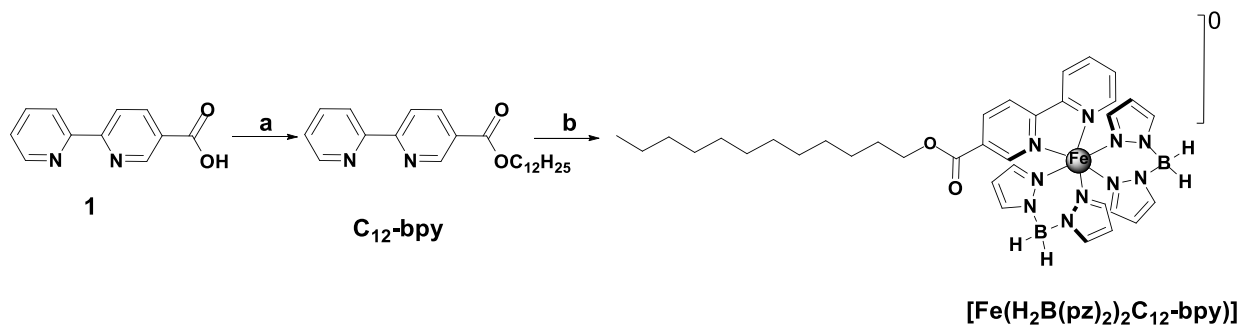

**Scheme S1.** Synthesis of [Fe(H$_2$B(pz)$_2$)$_2$(C$_{12}$-bpy)]. Key: (a) DCC/DMAP, DCM, RT and (b) Fe(ClO$_4$)$_2$·6H$_2$O, dihydro(bispyrazolyl)borate and C$_{12}$-bpy, MeOH/DCM.

Compound **1** was synthesized according to the literature procedure.[1]

**Synthesis of C$_{12}$-bpy**

An oven dried 50 mL Schlenk flask with a stir bar was charged with [2, 2'] Bipyridinyl-5-carboxylic acid (**1**) (0.29 g, 1.34 mmol) and 1-dodecanol (0.26 g, 1.34 mmol), and the solids were placed under argon atmosphere. To this, anhydrous CH$_2$Cl$_2$ (20 mL) was added and the mixture cooled to 0 °C in an Ice-Water bath. DCC (0.276 g, 1.34 mmol) and DMAP (0.163 g, 1.34 mmol) were then added to the stirring mixture as solids; the reaction mixture was slowly allowed to attain RT and stirred for 48 h followed by filtration to remove dicyclohexyl urea. Purification was accomplished by silica gel column chromatography using CH$_2$Cl$_2$ as an eluent. Yield: 0.185 g (37%). $^1$H NMR (300 MHz, CDCl$_3$, 300 K, TMS): $\delta$ = 9.1 (s, 1H), 8.5 (d, 8.1 Hz, 2H), 8.4 (dd, 8.4 Hz, 1H), 8.3 (d, 9.3 Hz, 1H), 8.2 (m, 4H), 8 (m, 3H), 7.9 (m, 2H), 7.8 (m, 1H), 4.5 (t, 6.6 Hz, 2H), 3.4 (t, 7.2 Hz, 2H) and 2.1 (m, 4H). $^{13}$C NMR (75 MHz, CDCl$_3$, 300 K, TMS): 28, 28.6, 33, 65, 120, 122, 123, 123.3, 124.3, 124.6, 124.8, 124.9, 125, 125.1, 126, 127.2, 127.3, 127.5, 128.6, 130, 131, 131.4, 136, 137, 134, 149, 150, 155, 159 and 165.3. Elemental Analysis: Calc. For: (C$_{23}$H$_{32}$N$_2$O$_2$): C, 74.96; H, 8.75; N, 7.6; Found: C, 74.87; H, 8.72; N, 7.6.

**Synthesis of [Fe(H$_2$B(pz)$_2$)$_2$(C$_{12}$-bpy)]**

To a solution of iron (II) perchlorate hydrate (0.073 g, 0.2 mmol) in 5 ml of methanol, potassium dihydro(bispyrazolyl)borate (0.075 g, 0.4 mmol) was added and the mixture was stirred for 15 mins followed by filtration to remove the precipitated KClO$_4$. A solution of C$_{12}$-bpy (0.074 g, 0.2 mmol) dissolved in 1 ml of dichloromethane solvent was added dropwise to the filtrate leading to the formation of greenish precipitate upon continuous stirring for 4 h under argon protection. The precipitate was filtered, washed with methanol and dried in a vacuum oven for 6 h at 50 °C to yield the title complex as a pale green powder.

Yield: 56 mg (38%). Elemental Analysis of the pale green powder: Calc. for [Fe(H$_2$B(pz)$_2$)$_2$(C$_{12}$-bpy)]: (C$_{36}$H$_{48}$B$_2$FeN$_{10}$O$_2$·1H$_2$O) C, 57.78; H, 6.73; N, 18.72; Found: C, 57.37; H, 6.57; N, 18.9.

**Sublimation of [Fe(H$_2$B(pz)$_2$)$_2$(C$_{12}$-bpy)]**

To verify the sublimability of the [Fe(H$_2$B(pz)$_2$)$_2$(C$_{12}$-bpy)] thermogravimetric analysis (TGA) was performed at an ambient pressure and argon atmosphere using an SDT Q600 apparatus from TA Instruments at a scanning rate of 5 °C min$^{-1}$. The result presented in Fig. S1 implies the onset of the sublimation at temperature $T_{sublim}$ ~ 150 °C, what was later used for fabrication of a 10 nm thick sample discussed in the main text.

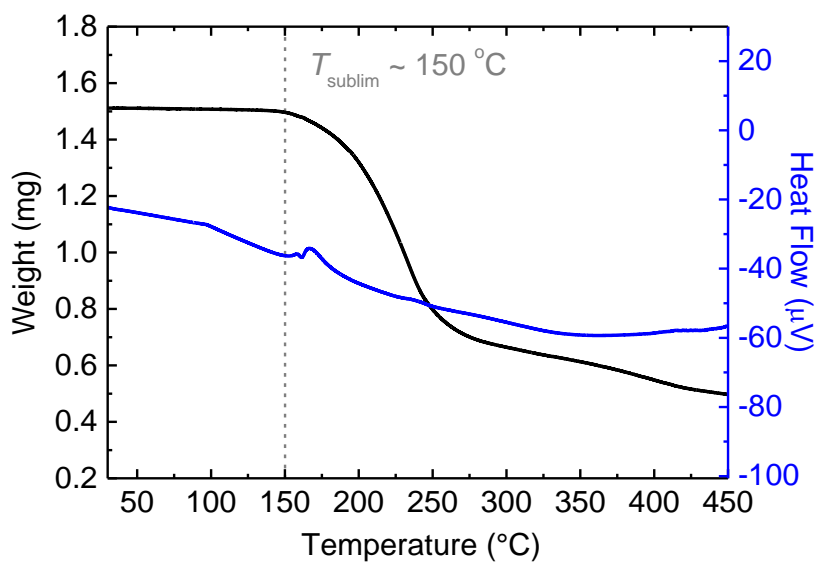

**Figure S1.** TGA of [Fe(H$_2$B(pz)$_2$)$_2$(C$_{12}$-bpy)]. The measurement was performed at an ambient pressure and argon atmosphere.

**Ultraviolet-visible spectroscopy**

To confirm the presence of [Fe(H$_2$B(pz)$_2$)$_2$(C$_{12}$-bpy)] molecules and its structural integrity on the quartz substrate after sublimation, ultraviolet-visible (UV-Vis) spectroscopy measurements were performed employing Varian Cary 5000 double-beam UV-vis-NIR spectrometer. A comparison between absorption spectrum of the [Fe(H$_2$B(pz)$_2$)$_2$(C$_{12}$-bpy)] complex in solution (cf. Fig. S2a) and thin complex film on quartz substrate revealed similar spectral features (cf. Fig. S2b) indicating the structural integrity of the complex in the sublimed film.

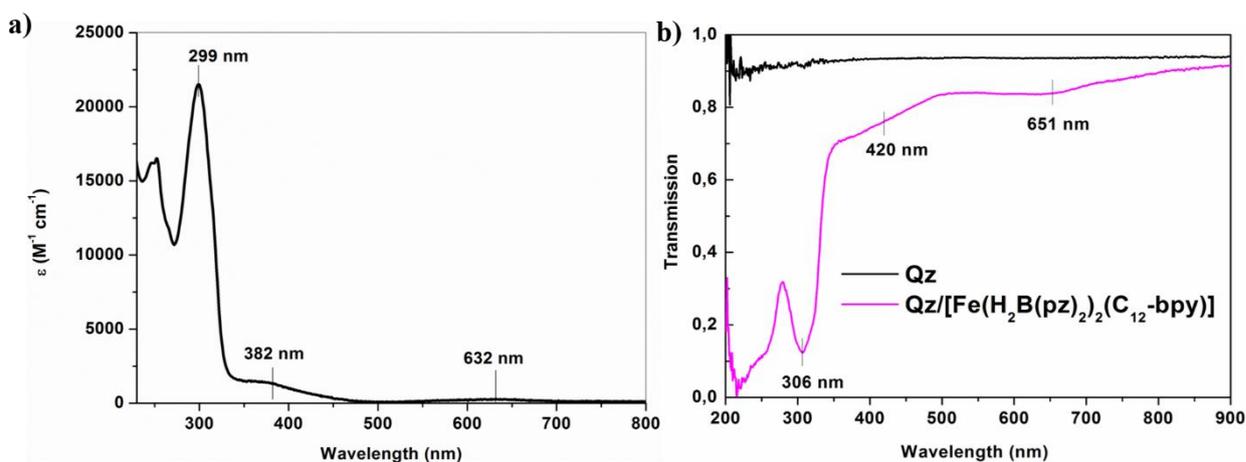

**Figure S2.** UV-Vis absorption and transmission spectra of [Fe(H$_2$B(pz)$_2$)$_2$(C$_{12}$-bpy)] (a) in chloroform solution and (b) as a film (specify the thickness) on quartz (Qz) glass substrate.

The peaks present at ca. 632 nm and 651 nm for the complex in solution and thin film state clearly indicate the HS state of the complex at ambient temperature in line with the experimental data presented in this manuscript. The bathochromic shift in the absorption maxima upon moving from solution to thin film state is tentatively attributed to strong intermolecular interactions between the complex entities.

**Variable temperature X-ray absorption measurements**

Variable temperature x-ray absorption measurements at Fe $L_{2,3}$ edges (Fig. S3) were performed to get insights into the SCO characteristics of the complex in powder and thin film state. The microcrystalline complex powder which also served as a reference, enabling comparison between bulk and thin film state, was pressed into indium foil. The spectra were recorded in total electron yield mode in an applied magnetic field of 0.1 T to facilitate the extraction of the electrons, at normal incidence and with linear x-ray polarization in the sample plane.

The XAS spectra at Fe $L_3$ edge of the thin film sample showed a single intersection point for all the curves at ~ 707.6 eV resulting from good uniformity of the film (Fig. S3b). On the other hand, deviation of the curves at this photon energy observed for the powder sample (Fig. S3a) could be attributed to the changes in the amount of probed matter due to the slight displacement of the x-ray beam position on the sample while varying the temperature.

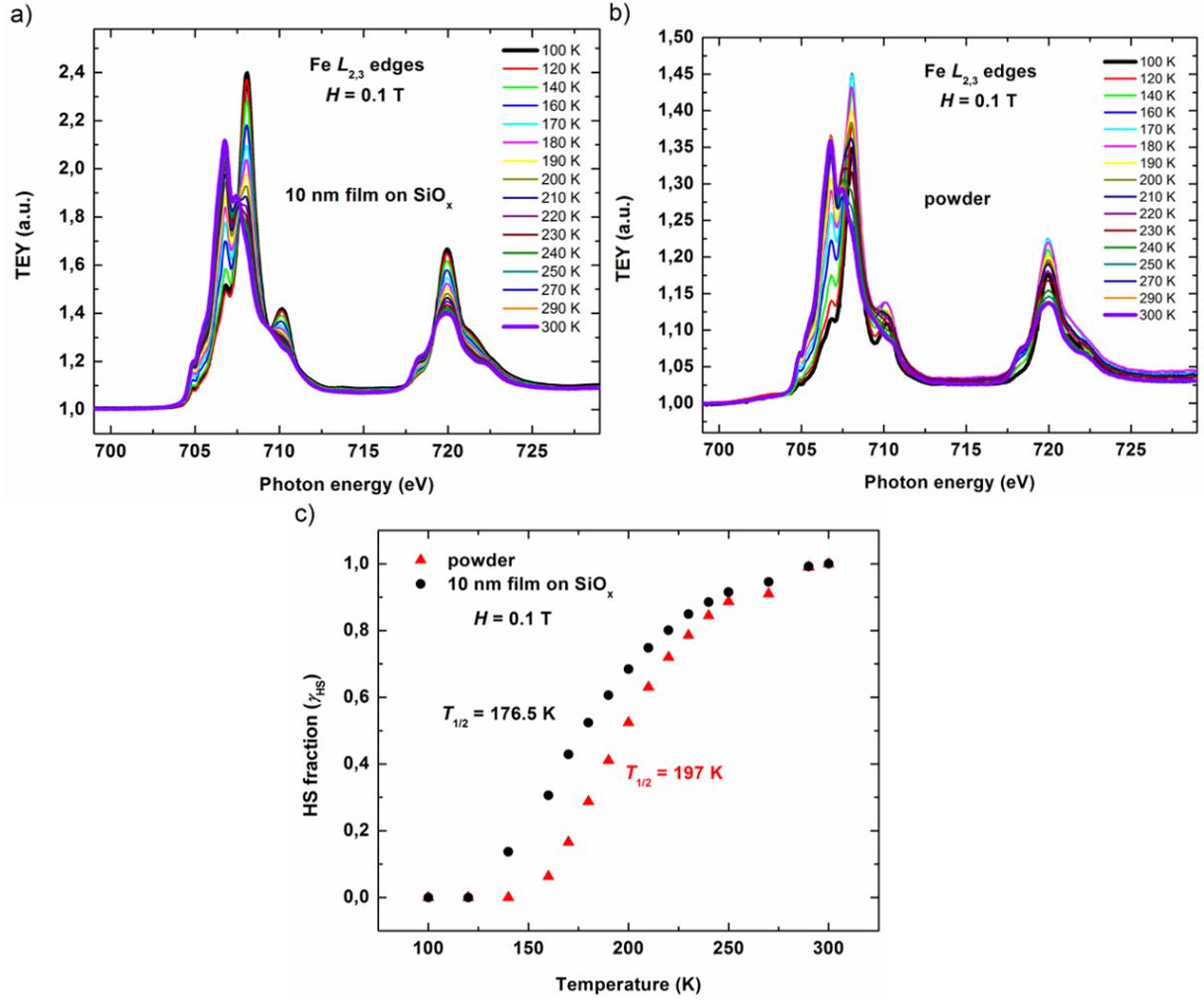

**Figure S3.** Variable temperature XAS at the Fe $L_{2,3}$ edges of (a) polycrystalline powder of [Fe(H$_2$B(pz)$_2$)$_2$(C$_{12}$-bpy)] complex, (b) 10 nm [Fe(H$_2$B(pz)$_2$)$_2$(C$_{12}$-bpy)] film on SiO$_x$, and (c) plot of the high spin fraction ($\gamma_{HS}$) vs. temperature for the powder and thin film samples of [Fe(H$_2$B(pz)$_2$)$_2$(C$_{12}$-bpy)] indicating nearly complete spin state switching of both powder and thin film samples with approximately 20 K reduction of the $T_{1/2}$ value for the thin film sample.

From the pre-edge normalized spectra we extracted high spin proportion for a given temperature based on the spectral intensity ratio of the features at ~ 706.8 eV and ~ 708 eV typical for HS $t_{2g}^4 e_g^2$ configuration (S = 2) and LS $t_{2g}^6 e_g^0$ (S = 0), respectively, according to the formula:

$$\frac{f(L_{3a})_x}{f(L_{3b})_x} = \frac{\gamma_{HS} f(L_{3a})_{HS} + (1-\gamma_{HS}) f(L_{3a})_{LS}}{\gamma_{HS} f(L_{3b})_{HS} + (1-\gamma_{HS}) f(L_{3b})_{LS}}$$

where $\gamma_{HS}$ is the high spin proportion, $f(L_{3a})$ and $f(L_{3b})$ are spectral intensities of the Fe $L_3$ edge feature at ~ 706.8 eV and ~ 708.1 eV (see Fig. 2 dashed vertical lines), respectively, and the indices x, HS and LS correspond to a mixed state spectrum, pure high spin (300 K) and pure low spin (100 K) references. The resulting curves of $\gamma_{HS}$ as a function of temperature for the film and powder sample are presented in Fig. S3.